\documentclass[graybox]{svmult}

\usepackage{mathptmx}       % selects Times Roman as basic font
\usepackage{helvet}         % selects Helvetica as sans-serif font
\usepackage{courier}        % selects Courier as typewriter font
\usepackage{type1cm}        % activate if the above 3 fonts are
\usepackage{makeidx}         % allows index generation
\usepackage{graphicx}        % standard LaTeX graphics tool
\usepackage{multicol}        % used for the two-column index
\usepackage[bottom]{footmisc}% places footnotes at page bottom

\usepackage{bm}
\usepackage{latexsym}
\usepackage{amsfonts,amssymb,amsmath}

\makeindex            

\begin{document}
\title*{What do detectors detect?}
\author{L.~Sriramkumar}
\institute{L.~Sriramkumar 
\at Department of Physics, Indian Institute of Technology Madras, 
Chennai 600036, India\\ 
\email{sriram@physics.iitm.ac.in}}
\maketitle
\abstract{By a detector, one has in mind a point particle with 
internal energy levels, which when set in motion on a generic 
trajectory can get excited due to its interaction with a 
quantum field. 
Detectors have often been considered as a helpful tool to 
understand the concept of a particle in a curved spacetime. 
Specifically, they have been used extensively to investigate 
the thermal effects that arise in the presence of horizons. 
In this article, I review the concept of detectors and discuss 
their response when they are coupled linearly as well as 
non-linearly to a quantum scalar field in different situations. 
In particular, I discuss as to how the response of detectors does 
not necessarily reflect the particle content of the quantum field. 
I also describe an interesting `inversion of statistics' that 
occurs in odd spacetime dimensions for `odd couplings', 
{\it i.e.}\/~the response of a uniformly accelerating detector 
is characterized by a Fermi-Dirac distribution even when it is 
interacting with a scalar field. 
Moreover, by coupling the detector to a quantum field that is 
governed by a modified dispersion relation arising supposedly 
due to quantum gravitational effects, I examine the possible 
Planck scale modifications to the response of a rotating 
detector in flat spacetime. 
Lastly, I discuss as to why detectors that are switched on for 
a finite period of time need to be turned on smoothly in order 
to have a meaningful response.}

%%%%%%%%%%%%%%%%%%%%%%%%%%%%%%%%%%%%%%%%%%%%%%%%%%%%%%%%%%%%%%%%%%%%%%%%%%%%%%%%

\section{Introduction}

The vacuum state of a quantum field develops a non-trivial structure 
in the presence of a strong classical electromagnetic or gravitational 
background.
This effect essentially manifests itself as two types of physical 
phenomena:~polarization of the vacuum and production of pairs of
particles corresponding to the quantum field.
Apart from these two effects, there is another feature that one encounters 
in a gravitational background:~the definition of the vacuum does not prove
to be generally covariant. 
In other words, the concept of a particle turns out to be, in general, 
dependent on the choice of coordinates.
(For a detailed discussion on these different aspects of quantum field theory 
in strong electromagnetic and gravitational fields, see the following 
texts~\cite{texts-qft-emb,texts-qft-cs} and reviews~\cite{qft-cb-reviews}.)
A classic example of vacuum polarization is the Casimir 
effect~\cite{casimir-1948}.
The Schwinger effect~\cite{heisenberg-1936,schwinger-1951}, {\it viz.}\/~pair 
creation by strong electric fields, and Hawking radiation from collapsing black 
holes are the most famous examples of particle production~\cite{hawking-1975}.
The coordinate dependence of the particle concept that arises in a 
gravitational background is well illustrated by the flat spacetime
example wherein the vacuum defined in the frame of a uniformly 
accelerating observer (often referred to as the Rindler vacuum) turns 
out to be inequivalent to the conventional Minkowski 
vacuum~\cite{fulling-1973}.
Similar issues are encountered when the behavior of quantum fields are 
studied in curved spacetimes.
Needless to say, concepts such as vacuum and particle need to be 
unambiguously defined in order to determine the extent of vacuum 
polarization or particle production occurring in a curved spacetime.

\par

It is in such a situation that the concept of a detector was
initially introduced in the literature~\cite{unruh-1976,dewitt-1979}.
The motivation behind the idea of detectors was to provide an operational 
definition for the concept of a particle in a curved spacetime. 
After all, `particles are what the particle detectors detect'~\cite{davies-1984}. 
With this goal in mind, the response of different types of detectors
have been studied in a variety of situations over the last three to 
four decades (in fact, there is an enormous amount of literature on
the topic; for an incomplete list of early efforts in this direction, 
see Refs.~\cite{letaw-1981a,detectors-general,detectors-odd-std,
detectors-cst,detectors-nlc,detectors-ft,sriram-1996,davies-1996,
suzuki-1997,sriram-1999,sriram-2002a,sriram-2002b,sriram-2003} and, 
for more recent work, see, for example, Refs.~\cite{detectors-recent,
detectors-nua,gutti-2011,psm-ue}).
But, what do these detectors actually detect? 
In particular, do their responses reflect the particle content of 
the field as it was originally desired?
In this article, apart from attempting to address such questions with
the help of a few specific examples, I shall also discuss a couple of 
interesting phenomena associated with detectors, including possible 
Planck scale effects. 
I should mention here that this article is essentially a review based 
on my earlier efforts in these directions (see Refs.~\cite{sriram-1996,
sriram-1999,sriram-2002a,sriram-2002b,gutti-2011}).
 
\par

An outline of the contents of this article is as follows.
In the following section, I shall discuss the response of non-inertial
Unruh-DeWitt detectors (which are linearly coupled to the quantum
field) in flat spacetime.
Specifically, I shall focus on the response of uniformly accelerating 
and rotating detectors.
I shall also compare the response of detectors in different situations
with the results from more formal methods---such as the Bogolubov 
transformations and the effective Lagrangian approach---that probe the 
vacuum structure of the quantum field.
Such an exercise helps us understand the conditions under which the 
detectors respond.  
In Sec.~\ref{sec:os-od-foc}, I shall consider the response of detectors
that are coupled non-linearly to a quantum scalar field.
Interestingly, I shall show that, in odd spacetime dimensions, the 
response of the detectors exhibit an `inversion of statistics' when
they are coupled to an odd power of the quantum field.
In Sec.~\ref{sec:pse}, I shall consider possible Planck scale 
effects on the response of a rotating detector in flat spacetime.
Assuming that the Planck scale effects modify the dispersion relation
governing a quantum field, I shall study the response of a rotating
Unruh-DeWitt detector that is coupled to such a quantum scalar field.
I shall illustrate that, while super-luminal dispersion relations 
hardly affect the response of the detector, 
sub-luminal dispersion relations alter their response considerably.
In Sec.~\ref{sec:ftd}, I shall consider Unruh-DeWitt detectors that 
are switched on for a finite period of time and show that divergences 
can arise in the response of the detector if it is turned on abruptly.
Lastly, I conclude in Sec.~\ref{sec:s} with a brief summary.   

\par

A few words on my conventions and notations are in order before I 
proceed.
I shall adopt natural units such that $\hbar=c=1$ and, for convenience, 
denote the trajectory of the detector $x^{\mu}(\tau)$ as ${\tilde x}(\tau)$,
where $\tau$ is the proper time in the frame of the detector.
In Sec.~\ref{sec:os-od-foc}, I shall consider the response of 
non-linearly coupled detectors in arbitrary spacetime dimensions.
In all the other sections, I shall restrict myself to working in 
$(3+1)$-spacetime dimensions.

%%%%%%%%%%%%%%%%%%%%%%%%%%%%%%%%%%%%%%%%%%%%%%%%%%%%%%%%%%%%%%%%%%%%%%%%%%%%%%%%

\section{Response of the Unruh-DeWitt detector in flat spacetime}

A detector is an idealized point like object whose motion 
is described by a classical worldline, but which nevertheless
possesses internal energy levels.
Such detectors are basically described by the interaction
Lagrangian for the coupling between the degrees of freedom
of the detector and the quantum field.
The simplest of the different possible detectors is the 
detector due to Unruh and DeWitt~\cite{unruh-1976,dewitt-1979}. 
Consider a Unruh-DeWitt detector that is moving along a 
trajectory ${\tilde x}(\tau)$, where $\tau$ is the proper 
time in the frame of the detector.
The interaction of the Unruh-DeWitt detector with a canonical, real 
scalar field~$\phi$ is described by the interaction Lagrangian
\begin{equation}
{\cal L}_{\rm int}[\phi({\tilde x})]
= {\bar c}\, m(\tau)\, \phi\left[{\tilde x}(\tau)\right],
\label{eqn:lint}
\end{equation}
where ${\bar c}$ is a small coupling constant and $m$ is the 
detector's monopole moment.
Let us assume that the quantum field ${\hat \phi}$ is 
initially in the vacuum state $\vert 0\rangle$ and the 
detector is in its ground state~$\vert E_0\rangle$ 
corresponding to an energy eigen value~$E_0$.
Then, up to the first order in perturbation theory, the 
amplitude of transition of the Unruh-DeWitt detector to 
an excited state~$\vert E_1\rangle$,  corresponding 
to an energy eigen value~$E_1~(>~E_0)$, is 
described by the integral~\cite{texts-qft-cs}
\begin{equation}
A(E) = M\, \int\limits_{-\infty}^{\infty} {\rm d}\tau\; 
{\rm e}^{i\, E\,\tau}\, 
\langle\psi\vert{\hat \phi}[{\tilde x}(\tau)]\vert 0\rangle,
\label{eq:ta}
\end{equation}
where $M= i\,{\bar c}\, \langle E_1\vert m(0) \vert E_{0}\rangle$, 
$E=E_1-E_0>0$ and $\vert \psi\rangle$ is the state of the 
quantum scalar field {\it after}\/ its interaction with the 
detector.
Note that the quantity $M$ depends only on the internal structure 
of the detector, and not on its motion. 
Therefore, as is often done, I shall drop the quantity hereafter.
The transition probability of the detector to all possible 
final states $\vert \psi\rangle$ of the quantum field is 
given by
\begin{equation}
P(E) = \sum_{\vert\psi\rangle} \vert A(E)\vert^2
= \int\limits_{-\infty}^\infty {\rm d}\tau\, 
\int\limits_{-\infty}^\infty {\rm d}\tau'\, 
{\rm e}^{-i\, E\,(\tau-\tau')}\, 
G^{+}\left[{\tilde x}(\tau), {\tilde x}(\tau')\right],\label{eqn:tp}
\end{equation}
where $G^{+}\left[{\tilde x}(\tau), {\tilde x}(\tau')\right]$ 
is the Wightman function defined as
\begin{equation}
G^{+}\left[{\tilde x}(\tau), {\tilde x}(\tau')\right]
=\langle 0 \vert 
{\hat \phi}\left[{\tilde x}(\tau)\right]\,
{\hat \phi}\left[{\tilde x}(\tau')\right]
\vert 0 \rangle.\label{eqn:wfn}
\end{equation}

\par

When the Wightman function is invariant under time translations in the 
frame of the detector---as it can occur, for example, in cases wherein 
the detector is moving along the integral curves of time-like Killing 
vector fields~\cite{letaw-1981a,sriram-2002a}---I have
\begin{equation}
G^{+}\left[{\tilde x}(\tau), {\tilde x}(\tau')\right]
= G^{+}(\tau-\tau').
\end{equation}
In such situations, the transition probability of the detector 
simplifies to
\begin{equation}
P(E)= \lim_{T\to \infty}\;\int\limits_{-T}^{T}\; \frac{{\rm d}v}{2}\, 
\int\limits_{-\infty}^{\infty}\, {\rm d}u\; 
{\rm e}^{-i\,E\, u}\; G^{+}(u),\label{eq:tp} 
\end{equation}
where 
\begin{equation}
u=\tau-\tau'\quad{\rm and}\quad v=\tau+\tau'.\label{eq:uv}
\end{equation}
The above expression then allows one to define the transition probability 
{\it rate}\/ of the detector to be~\cite{texts-qft-cs}
\begin{eqnarray}
R(E) 
= \lim_{T\to \infty}\; \frac{P(E)}{T}
= \int\limits_{-\infty}^\infty\, {\rm d}u\; 
{\rm e}^{-i\, E\, u}\; G^{+}(u).\label{eq:tpr}
\end{eqnarray}
For the case of the canonical, massless scalar field, in $(3+1)$-spacetime
dimensions, the Wightman function $G^{+}\left({\tilde x}, {\tilde x}'\right)$ 
in the Minkowski vacuum is given by~\cite{texts-qft-cs}
\begin{equation}
G^+\left({\tilde x},{\tilde x}'\right)
=-\frac{1}{4\, \pi^2}\,
\left[\frac{1}{(t-t'-i\,\epsilon)^2
-\left({\bm x}-{\bm x}'\right)^2}\right],\label{eq:wfn-mv}
\end{equation}
where $\epsilon\to 0^{+}$ and $(t,{\bm x})$ denote the Minkowski 
coordinates.
Given a trajectory ${\tilde x}(\tau)$, the response of the detector is 
obtained by substituting the trajectory in this Wightman function and 
evaluating the transition probability rate~(\ref{eq:tpr}).
For example, it is straightforward to show that the response of a detector 
that is moving on an inertial trajectory in the Minkowski vacuum vanishes 
identically.
I had mentioned above that the quantization of a field proves to be 
inequivalent in the inertial and the uniformly accelerating frames in 
flat spacetime.
Due to this reason, it seems worthwhile to examine the behavior of 
non-inertial detectors.
In the next sub-section, I shall consider the response of uniformly
accelerating as well as rotating detectors in flat spacetime.
 
%%%%%%%%%%%%%%%%%%%%%%%%%%%%%%%%%%%%%%%%%%%%%%%%%%%%%%%%%%%%%%%%%%%%%%%%%%%%%%%%

\subsection{Response of accelerating and rotating detectors}

As is commonly known, there are ten independent time-like Killing 
vector fields in flat spacetime.
These Killing vector fields correspond to three types of symmetries, 
{\it viz.}\/ translations, rotations and boosts.
Different types of non-inertial trajectories can be generated by 
considering the integral curves of various linear combinations 
of these Killing vector fields~\cite{letaw-1981a,sriram-2002a}.
Amongst the  trajectories that are possible, there exist two 
trajectories which have attracted considerable attention in the 
literature.
They correspond to uniformly accelerating and rotating trajectories.
In what follows, I shall consider the response of the Unruh-DeWitt
detector moving along these trajectories.
 
%%%%%%%%%%%%%%%%%%%%%%%%%%%%%%%%%%%%%%%%%%%%%%%%%%%%%%%%%%%%%%%%%%%%%%%%%%%%%%%%

\subsubsection{Uniformly accelerated motion}\label{subsubsec:rind}

The trajectory of a uniformly accelerated observer moving along the 
$x$-axis is given by
\begin{equation}
{\tilde x}(\tau)
=g^{-1}\,\left[{\rm sinh}\,(g\,\tau),\, 
{\rm cosh}\,(g\,\tau),\, 0,\,0\right],\label{eq:uat}
\end{equation}
where $g$ denotes the proper acceleration.
The coordinates associated with the frame of such an observer are 
known as the Rindler coordinates~\cite{rindler-1966}.
The Wightman function in the frame of the uniformly accelerating 
observer is obtained by substituting the above trajectory in  
Eq.~(\ref{eq:wfn-mv}).
It is given by 
\begin{eqnarray}
G^{+}(u)
=\frac{-1}{16\,\pi^2}\,
\frac{g^2}{{\rm sinh}^2\left[\left(g\,u/2\right)-i\,\epsilon\right]}
=\frac{-1}{4\,\pi^2}\,
\sum_{n=-\infty}^{\infty}\frac{1}{\left(u-i\,\epsilon
+2\,\pi\, i\,n/g\right)^{2}},\label{eq:wfn-mv-uat}
\end{eqnarray}
where, recall that, $u=\tau-\tau'$. 
The resulting transition probability rate can be easily evaluated 
to be~\cite{unruh-1976,dewitt-1979}
\begin{equation}
R(E)=\frac{1}{2\,\pi}\,
\frac{E}{{\rm e}^{2\,\pi\, E/g}-1},
\label{eq:r-uad}
\end{equation}
which is a thermal spectrum corresponding to the temperature~$T=g/(2\,\pi)$.
This thermal response is the famous Unruh effect (for a detailed discussion, 
see, for instance, Ref.~\cite{crispino-2008}).

%%%%%%%%%%%%%%%%%%%%%%%%%%%%%%%%%%%%%%%%%%%%%%%%%%%%%%%%%%%%%%%%%%%%%%%%%%%%%%%%

\subsubsection{Rotational motion}\label{subsubsec:rot} 

\begin{figure}[!t]
\begin{center}
\resizebox{285pt}{190pt}{\includegraphics{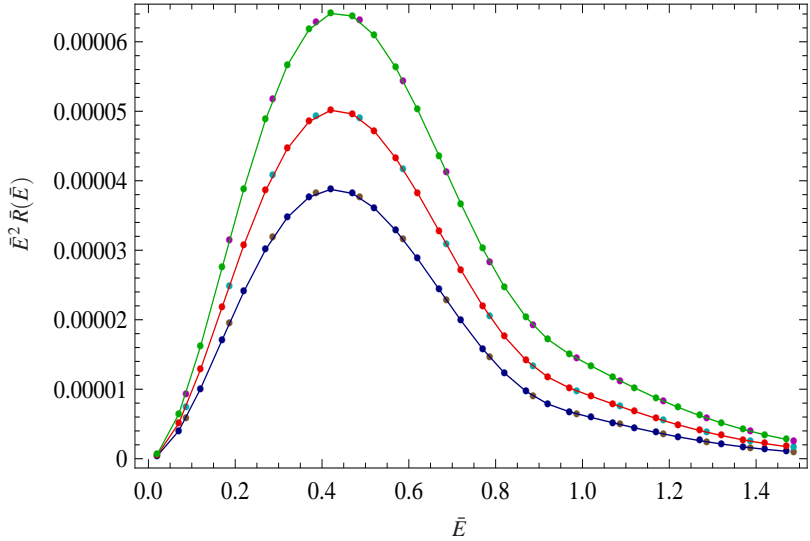}}
\vskip -125 pt \hskip -310 pt 
\rotatebox{90}{${\bar E}^2\, {\bar R}({\bar E})$}
\vskip 98 pt
\hskip 22 pt ${\bar E}$ 
\caption{The transition probability rate of the rotating Unruh-DeWitt 
detector that is coupled to the conventional, massless scalar field.
The dots and the curves that simply link them represent numerical 
results arrived at from the computation of the integral~(\ref{eq:tpr}) 
along the rotating trajectory.
The curves correspond to the following three values of the quantity 
$\sigma\, \Omega=0.325$ (in blue), $0.350$ (in red) and $0.375$ (in 
green). 
The dots of an alternate color that appear on the curves denote the 
numerical results that have been obtained by another method which I 
shall describe below [they actually correspond to the 
sum~(\ref{eq:tpr-sc-rt-smfv})].
Clearly, the results from the two methods match very well.}
\label{fig:rd-sc}
\end{center}
\end{figure}

Let us now turn to the case of the rotating detector.
The trajectory of the rotating detector can be expressed in terms of 
the proper time $\tau$ as follows~\cite{letaw-1981a,gutti-2011}:
\begin{equation}
{\tilde x}(\tau)
=\left[\gamma\,\tau,\,\sigma\, \cos\,(\gamma\,\Omega\,\tau),\,
\sigma\, \sin\, (\gamma\,\Omega\,\tau),\,0\right],\label{eq:rt}
\end{equation}
where the constants $\sigma$ and $\Omega$ denote the radius of the circular 
path along which the detector is moving and the angular velocity of the 
detector, respectively.
The quantity $\gamma=\left[1-(\sigma\, \Omega)^2\right]^{-1/2}$ is the Lorentz 
factor that relates the Minkowski time to the proper time in the frame 
of the detector.
The Wightman function along the rotating trajectory can be obtained to be
\begin{equation}
G^+(u)
=-\frac{1}{4\,\pi^2}\,
\left(\frac{1}{\gamma^2\, (u-i\,\epsilon)^2
-4\,\sigma^2\,\sin^2\,(\gamma\, \Omega\, u/2)}\right).
\label{eq:mgfn-sc-rt}
\end{equation}
However, unfortunately, it does not seem to be possible to evaluate the 
corresponding transition probability rate $R(E)$ analytically. 
I have arrived at the response of the rotating detector by substituting the 
Wightman function~(\ref{eq:mgfn-sc-rt}) in the expression~(\ref{eq:tpr}), 
and numerically computing the integral involved.
If I define the dimensionless energy to be ${\bar E}=E/(\gamma\, \Omega)$, 
I find that the dimensionless transition probability rate 
${\bar R}({\bar E})=\sigma\, R({\bar E})$ of the 
detector depends only on the dimensionless quantity $\sigma\, \Omega$ 
that describes the linear velocity of the detector.
In Fig.~\ref{fig:rd-sc}, I have plotted the transition probability rate 
of the detector for three different values of the quantity $\sigma\, 
\Omega$~\cite{letaw-1981a}.
%%%%%%%%%%%%%%%%%%%%%%%%%%%%%%%%%%%%%%%%%%%%%%%%%%%%%%%%%%%%%%%%%%%%%%%%%%%%%%%%
%%%%%%%%%%%%%%%%%%%%%%%%%%%%%%%%%%%%%%%%%%%%%%%%%%%%%%%%%%%%%%%%%%%%%%%%%%%%%%%%
I should mention here that, in order to check the accuracy of the 
numerical procedure that I have used to evaluate the 
integral~(\ref{eq:tpr}) for the rotating trajectory, I have compared 
the results from the numerical code with the analytical one [{\it viz.}\/
Eq.~(\ref{eq:r-uad})] that is available for the case of the uniformly 
accelerated detector. 
This comparison clearly indicates that the numerical procedure I have
adopted to evaluate the integral~(\ref{eq:tpr}) is quite 
accurate~\cite{gutti-2011}.

\par

In the discussion above, I had arrived at the response of the rotating 
detector by evaluating the Fourier transform of the Wightman function 
with respect to the differential proper time $u$ in the frame of the 
detector.
In this case, evidently, I had first summed over the normal modes (to 
arrive at the Wightman function) before evaluating the integral over the 
differential proper time.
I shall now rederive the result by changing the order of these procedures.
I shall express the Wightman function as a sum over the normal modes and 
first evaluate the integral over the differential proper time before 
computing the sum.
This method proves to be helpful later when I shall consider the Planck scale
effects on the rotating detector.
As I shall illustrate, the method can be easily extended to cases wherein 
the scalar field is described by a modified dispersion relation.

\par

I shall start by working in the cylindrical polar coordinates, say, 
$(t,\rho,\theta,z)$, instead of the cartesian coordinates, since they 
prove to be more convenient.
In terms of the cylindrical coordinates, the trajectory~(\ref{eq:rt}) of 
the rotating detector can be written in terms of the proper time $\tau$
as follows:
\begin{equation}
{\tilde x}(\tau)
=\left(\gamma\, \tau,\,\sigma,\,\gamma\, \Omega\, \tau,\, 0\right).
\label{eq:rt-cyc}
\end{equation}
Using well established properties of the Bessel functions, it is
straightforward to show that, along the trajectory of the rotating 
detector, the standard Minkowski Wightman function~(\ref{eq:wfn-mv}) 
can be written as 
\begin{equation}
G^{+}(u)
=\sum^{\infty}_{m=-\infty}\,
\int\limits^\infty_0\frac{{\rm d}q\, q}{\left(2\,\pi\right)^2}\,
\int\limits^\infty_{-\infty}\frac{{\rm d}k_z}{\left(2\,\omega\right)}\,
J_{m}^{2}(q\,\sigma)\; 
{\rm e}^{-i\,\gamma\,\left(\omega-m\,\Omega\right)\, u},
\label{eq:mgfn-rt-ms}
\end{equation} 
where $J_{m}(q\, \sigma)$ denote the Bessel functions of order $m$, with 
$\omega$ being given by
\begin{equation}
\omega=\left(q^{2}+k_{z}^{2}\right)^{1/2}.
\end{equation}
One can then immediately express the corresponding transition probability 
rate of the rotating detector as [cf.~Eq.~(\ref{eq:tpr})]
\begin{equation}
R(E)=\sum^{\infty}_{m=-\infty}\,
\int\limits^\infty_0\frac{{\rm d}q\, q}{2\,\pi}\;
\int\limits^\infty_{-\infty}\;\frac{{\rm d}k_z}{2\,\omega}\,
J_{m}^{2}(q\,\sigma)\; 
\delta^{(1)}\left[E+\gamma\, \left(\omega-m\,\Omega\right)\right].
\label{eq:tp-cc}
\end{equation}
Recall that, $E>0$, $\omega\ge 0$ (as is appropriate for positive frequency 
modes), and I have assumed that $\Omega$ is a positive definite quantity as
well.
Hence, the delta function in the above expression will be non-zero only when 
$m \ge {\bar E}$, where ${\bar E}=E/\left(\gamma\, \Omega\right)$ is the 
dimensionless energy.
Due to this reason, the response of the detector simplifies to
\begin{equation}
R({\bar E})
=\sum^{\infty}_{m\ge {\bar E}}\,
\int\limits^\infty_0\frac{{\rm d}q\, q}{2\,\pi}\;
\int\limits^\infty_{-\infty}\frac{{\rm d}k_z}{2\,\omega}\;
J_{m}^{2}(q\,\sigma)\; 
\left[\frac{\delta^{(1)}(k_z-\kappa_{z})}{\gamma\,
\left|\left({\rm d}\omega/{\rm d}k_z\right)\right|_{\kappa_z}}\right],
\label{eq:tpr-sc-rt-msev}
\end{equation}
where $\kappa_{z}$ are the two roots of $k_{z}$ from the following 
equation: 
\begin{equation}
\omega= \left(m-{\bar E}\right)\, \Omega.\label{eq:r-eq}
\end{equation}
The roots are given by
\begin{equation}
\kappa_z=\pm\left(\lambda^{2}-q^2\right)^{1/2},\label{eq:r-kzb-sc}
\end{equation}
where, for convenience, I have set 
\begin{equation}
\lambda={\bar \lambda}\, \Omega
=\left(m-{\bar E}\right)\, \Omega.\label{eq:delta}
\end{equation}
Since both the positive and negative roots of $\kappa_{z}$ contribute 
equally, the dimensionless transition probability rate of the rotating 
detector can be obtained to be
\begin{equation}
{\bar R}({\bar E}) = \sigma\, R({\bar E})
=\frac{\sigma}{2\,\pi\,\gamma}\,
\sum^{\infty}_{m\ge {\bar E}}\,
\int\limits_{0}^{\lambda}\; {\rm d}q\; q\;
\left[\frac{J_{m}^{2}(q\,\sigma)}{\left(\lambda^{2}-q^{2}\right)^{1/2}}\right],
\end{equation}
where I have set the upper limit on $q$ to be $\lambda$ as $\kappa_{z}$ 
is a real quantity [cf.~Eq.~(\ref{eq:r-kzb-sc})].
I find that the integral over~$q$ can be expressed in terms of the 
hypergeometric functions (see, for instance, Ref.~\cite{prudnikov-1986}).
Therefore, the transition probability rate of the rotating detector can
be written as 
\begin{eqnarray}
{\bar R}({\bar E})
&=&\frac{1}{2\,\pi\,\gamma}\
\sum^{\infty}_{m\ge {\bar E}}\,
\frac{\left(\sigma\,\Omega\, 
{\bar \lambda}\right)^{(2\,m+1)}}{\Gamma\left(2\,m+2\right)}\nonumber\\ 
& &\times\,_1F_2\left[m+(1/2);\,m+(3/2),2\,m+1;\,
-\left(\sigma\,\Omega\, {\bar \lambda}\right)^2\right],
\label{eq:tpr-sc-rt-smfv}
\end{eqnarray}
where $_1F_2\left(a;b,c;x\right)$ denotes the hypergeometric function, while 
$\Gamma(x)$ is the usual Gamma function.
Though it does not seem to be possible to arrive at a closed form expression 
for this sum, the sum converges very quickly, and hence proves to be easy 
to evaluate numerically.
In Fig.~\ref{fig:rd-sc}, I have plotted the numerical results for the
above sum for the same values of the linear velocity $\sigma\, \Omega$ 
for which I had plotted the results obtained from Fourier transforming
the Wightman function~(\ref{eq:mgfn-sc-rt}) along the rotating trajectory.
The figure clearly indicates that the results from the two different 
methods match each other rather well. 

%%%%%%%%%%%%%%%%%%%%%%%%%%%%%%%%%%%%%%%%%%%%%%%%%%%%%%%%%%%%%%%%%%%%%%%%%%%%%%%%

\subsection{Are detectors sensitive to the particle content
of the field?}\label{subsec:comparion}

In order to clearly understand as to what detectors detect, I shall
compare the response of detectors with the results from more conventional 
probes of the vacuum structure of the quantum fields, such as the 
approaches based on the Bogolubov transformations and the effective 
Lagrangian~\cite{sriram-2002a}.
However, before carrying out such a comparison, let me say a few words
briefly explaining these two other approaches.

\par

Consider a quantum field that can be decomposed in terms of two
complete sets of normal modes.
These two sets of modes can be related to each other through the 
Bogolubov transformations, which are essentially characterized by 
two coefficients often referred to as $\alpha$ and 
$\beta$~\cite{bogolubov-1958}.
Moreover, the particle content of the field is determined by the 
Bogolubov coefficient~$\beta$.
In a gravitational background, the Bogolubov transformations can 
either relate the modes of a quantum field at two different times 
in the same coordinate system or the modes in two different 
coordinate systems covering the same region of spacetime.
When the Bogolubov coefficient~$\beta$ is non-zero, in the latter 
context, such a result is normally interpreted as implying that 
the quantization in the two coordinate systems are 
inequivalent~\cite{fulling-1973}.
Whereas, in the former context, a non-zero~$\beta$ is attributed 
to the production of particles by the background gravitational 
field~\cite{qft-cb-reviews}.
Similarly, in an electromagnetic background, a non-zero~$\beta$ 
relating the modes of a quantum field at different times (in a 
particular gauge) implies that the background leads to pair 
creation~\cite{texts-qft-emb}.

\par

In the effective Lagrangian approach, one essentially integrates
out the degrees of freedom associated with the quantum field,
thereby arriving at an effective action describing the classical
background~\cite{heisenberg-1936,schwinger-1951}.
An imaginary part to the effective Lagrangian unambiguously 
suggests the decay of the quantum vacuum, {\it i.e.}\/ the production
of particles corresponding to the quantum field. 
The real part of the effective Lagrangian can be related
to the extent of polarization of the vacuum caused by the
classical background.
While the effective Lagrangian approach is powerful, since it 
involves computing a path integral, it often proves to be 
technically difficult to evaluate.

\par
 
In Tab.~\ref{tab:comparison}, to illustrate the conclusions I wish 
to draw about the response of detectors, I have tabulated the results 
one obtains in a handful of different situations.
I have listed whether the Bogolubov coefficient~$\beta$, the response
of the detector [or, more precisely, the transition probability $P(E)$]
and the real and the imaginary parts of the effective 
Lagrangian~${\cal L}_{\rm eff}$ are zero or non-vanishing in these 
contexts.
Apart from the results in the non-inertial frames in flat spacetime,
I have compared the results between the Casimir plates, and different
types of electromagnetic backgrounds.
%%%%%%%%%%%%%%%%%%%%%%%%%%%%%%%%%%%%%%%%%%%%%%%%%%%%%%%%%%%%%%%%%%%%%%%%%%%%%%%%
\begin{table}[!t]
\begin{center}
\begin{tabular}{|c|c|c|c|}
\hline
& Detector &  Bogolubov   &  Effective \\ 
& response & coefficient  & Lagrangian \\ 
& $P(E)$ & $\beta$  
& ${\rm Re.}~{\cal L}_{\rm eff} \quad
{\rm Im.}~{\cal L}_{\rm eff}$ \\
\hline
In inertial coordinates & $0$ &  $0$ & $\!\!\!\! 0 
\qquad\qquad\;\; 0$\\
\hline
In Rindler coordinates & $\ne 0$ & $\ne 0$ 
& $\!\!\!\! 0\qquad\qquad\;\; 0$\\
\hline
In rotating coordinates & $\ne 0$ & $0$ 
& $\!\!\!\! 0 \qquad\qquad\;\; 0$\\
\hline
Between Casimir plates & $0$ & $0$ 
& $\!\!\!\!\!\!\!\ne 0\qquad\quad\;\;\;\,0$\\
\hline
In a time-dependent & $\ne 0$ & $\ne 0$ 
& $\!\!\!\!\,\ne 0\qquad\quad\,\,\ne 0$\\
electric field & & &\\
\hline
In a time-independent & $\;\,\ne 0$
& $\ne 0$ & $\!\!\!\!\,\ne 0\qquad\quad\,\,\ne 0$\\
electric field & & &\\
\hline
In a time-independent & $0$ & $0$  
& $\!\!\!\!\!\!\!\ne 0\qquad\quad\;\;\;\,0$\\
magnetic field & & &\\
\hline
\end{tabular}
\end{center}
\caption{A comparison of the response of a detector with the results
from more formal probes of the vacuum structure of the quantum 
field---{\it viz.}\/ the Bogolubov transformations and the effective 
Lagrangian approaches---in a variety of situations.
Note that, in the case of the time-independent electric field background,
actually, the Bogolubov coefficient $\beta$ is trivially zero.
I refer here to particle production that can occur in such a background 
due to the phenomenon called Klein paradox~\cite{klein-p}.
I should also add that, in electromagnetic backgrounds, the coupling of 
the detector to the quantum field (say, a complex scalar field) has to
be intrinsically non-linear in order to preserve 
gauge-invariance~\cite{sriram-1999}.}
\label{tab:comparison}
\end{table}
%%%%%%%%%%%%%%%%%%%%%%%%%%%%%%%%%%%%%%%%%%%%%%%%%%%%%%%%%%%%%%%%%%%%%%%%%%%%%%%%

\par

Let me first consider the case of the non-inertial coordinates in flat
spacetime.
The Bogolubov coefficient~$\beta$ relating the Rindler modes and the 
Minkowski modes turns out to be non-zero and, in fact, the expectation 
value of the Rindler number operator in the Minkowski vacuum yields a
thermal spectrum as well~\cite{fulling-1973}. 
In contrast, in the rotating coordinates, while the Bogolubov 
coefficient~$\beta$ turns out to be zero~\cite{letaw-1981a}, 
as we have seen, the detector responds non-trivially.
Also, in both these cases, one can show that the effective Lagrangian 
vanishes identically---in fact, this is true even in the case of the 
Rindler coordinates, wherein the Bogolubov coefficient~$\beta$ proves 
to be non-zero~\cite{sriram-2002a}.
Evidently, the response of a detector can be non-zero even when the 
Bogolubov coefficient~$\beta$ and the effective Lagrangian vanish 
identically.
Clearly, the response of a detector does not necessarily reflect the 
particle content of the quantum field.

\par

Let me now turn to the response of the detector between Casimir plates
and in electromagnetic backgrounds.
It is well known that Casimir plates and a time-independent magnetic 
field lead to vacuum polarization, but not to particle production.  
One finds that an inertial detector does not respond in these two 
backgrounds.
In contrast, it is found that even an inertial detector responds in
an electric field background, whether time-dependent or otherwise.  
It is easy to argue that, in a time-dependent electric field, the 
evolving modes will excite the inertial detector~\cite{sriram-1999,
sriram-2002a}.
Whereas, in a time-independent electric field of sufficient strength, 
modes of positive norm that have negative frequencies (which lead to
the so-called Klein paradox and associated pair 
production~\cite{sriram-2002a,klein-p}, as is also reflected by the 
imaginary part of the effective Lagrangian~\cite{schwinger-1951}) 
are found to be responsible for a non-vanishing response of an inertial 
detector\footnote{In fact, it is such modes---{\it viz.}\/ those which 
have a positive norm but negative frequencies---that excite the 
rotating detector~\cite{sriram-2002a,letaw-1981b}.}.
These clearly suggest that, irrespective of the nature of its trajectory, 
a detector will respond whenever particle production takes place.
In that sense a detector {\it is}\/ sensitive to particle production.
Further, if one restricts the motion of the detector to inertial 
trajectories, then the effects due to non-inertial motion can be
avoided and, in such cases, the detector response will be non-zero 
{\it only}\/ when particle production takes place. 
However, unlike in flat spacetime or classical electromagnetic 
backgrounds, there exists no special frame of reference in a 
classical gravitational background and all coordinate systems 
have to be treated equivalently.  
This aspect of the detector proves to be a major constraint in
being able to utilize it to investigate the phenomenon of particle 
production in a curved spacetime~\cite{davies-1984,particles-cst}.

%%%%%%%%%%%%%%%%%%%%%%%%%%%%%%%%%%%%%%%%%%%%%%%%%%%%%%%%%%%%%%%%%%%%%%%%%%%%%%%%

\section{`Inversion of statistics' in odd dimensions}\label{sec:os-od-foc}

We had seen that the response of a uniformly accelerating monopole 
detector that is coupled to a quantized massless scalar field is 
characterized by a Planckian distribution when the field is 
assumed to be in the Minkowski vacuum [cf.~Eq.~(\ref{eq:r-uad})].
However, it has been noticed that this result is true only in 
even-dimensional flat spacetimes and it has been shown that a 
Fermi-Dirac factor (rather than a Bose-Einstein factor) appears 
in the response of the accelerated detector when the dimensionality 
of spacetime is odd~\cite{detectors-odd-std}. 
Recall that the Unruh-DeWitt detector is coupled {\it linearly}\/ to 
the quantum scalar field.
Over the years, motivated by different reasons, there have also been 
efforts in the literature to investigate the response of detectors 
that are coupled {\it non-linearly}\/ to the quantum 
field~\cite{detectors-nlc,suzuki-1997,sriram-1999}.
It will be interesting to examine whether the non-linearity 
of the coupling affects the result in odd-dimensional flat 
spacetimes that I mentioned above.

%%%%%%%%%%%%%%%%%%%%%%%%%%%%%%%%%%%%%%%%%%%%%%%%%%%%%%%%%%%%%%%%%%%%%%%%%%%%%%%%

\subsection{Response of non-linearly coupled detectors}

Consider a detector that is interacting with a real scalar field~$\phi$ 
through the non-linear interaction Lagrangian~\cite{suzuki-1997}
\begin{equation}
{\cal L}_{\rm int}[\phi({\tilde x})]
= {\bar c}\, m(\tau)\; \phi^n\left[{\tilde x}(\tau)\right],
\label{eqn:nlint}
\end{equation}
where ${\bar c}$, $m(\tau)$ and ${\tilde x}(\tau)$ are the same
quantities that we had encountered earlier in the context of the
Unruh-DeWitt detector.
The quantity $n$ is a positive integer that denotes the index of 
non-linearity of the coupling.
Let me assume that the quantum field ${\hat \phi}$ is initially 
in the vacuum state $\left\vert 0 \right\rangle$.
The transition amplitude of the non-linearly coupled detector 
from the ground to an excited state can be written as
\begin{equation}
A_n(E) = M\,
\int\limits_{-\infty}^{\infty} {\rm d}\tau\, {\rm e}^{i\, E\,\tau}\, 
\left\langle\psi\right\vert\, {\hat \phi}^n[{\tilde x}(\tau)]\,
\left\vert 0\right\rangle,\label{eq:ta-nlcd}
\end{equation}
where $\left\vert \psi \right\rangle$ is the final state of the field,
and $M$ and $E$ are defined in the same fashion as in the case of the 
Unruh-Dewitt detector. 

\par

It is important to notice that the transition amplitude~$A_{n}(E)$ above 
involves products of the quantum field~${\hat \phi}$ at the {\it same}\/ 
spacetime point.
Because of this reason, one will encounter divergences when evaluating 
this transition amplitude.
In order to avoid these divergences, I shall normal order the operators 
in the matrix element in the transition amplitude~$A_{n}(E)$ with 
respect to the Minkowski vacuum~\cite{suzuki-1997}.
In other words, rather than the expression~(\ref{eq:ta-nlcd}), I shall 
assume that the transition amplitude is instead given by
\begin{equation}
{\bar A}_{n}(E) 
=\int\limits_{-\infty}^{\infty} {\rm d}\tau\, {\rm e}^{i\, E\,\tau}\, 
\left\langle\psi\right\vert
:{\hat \phi}^n[{\tilde x}(\tau)]:\left\vert 0\right\rangle,
\label{eq:no-ta-nlcd}
\end{equation}
where the colons denote normal ordering with respect to the Minkowski vacuum.
Then, the transition probability of the detector to all 
possible final states $\left\vert \psi\right\rangle$ of 
the quantum field can be written as
\begin{equation}
P_{n}(E) 
=\sum_{\vert\psi\rangle}{\vert 
{\bar A}_{n}(E)\vert}^2
= \int\limits_{-\infty}^\infty {\rm d}\tau\, 
\int\limits_{-\infty}^\infty {\rm d}\tau'\, 
{\rm e}^{-i\,E\,(\tau-\tau')}\;
G_{n}\left[{\tilde x}(\tau), {\tilde x}(\tau')\right],
\label{eq:tp-nlcd}
\end{equation}
where $G_{n}\left[{\tilde x}(\tau), {\tilde x}(\tau')\right]$ 
is the $(2\,n)$-point function defined as
\begin{equation}
G_{n}\left[{\tilde x}(\tau), {\tilde x}(\tau')\right]
=\left\langle 0 \right\vert 
:{\hat \phi}^n\left[{\tilde x}(\tau)\right]:\,
:{\hat \phi}^n\left[{\tilde x}(\tau')\right]:
\left\vert 0 \right\rangle.\label{eqn:2npt-fn}
\end{equation}
In situations where the $(2\,n)$-point function 
$G_{n}\left[{\tilde x}(\tau), {\tilde x}(\tau')\right]$ 
is invariant under time translations in the frame of the detector, 
I can define a transition probability rate for the detector 
as follows:
\begin{equation}
R_{n}(E) 
= \int\limits_{-\infty}^\infty {\rm d}u\;
{\rm e}^{-i\, E\, u}\; G_{n}(u),\label{eq:tpr-nlcd}
\end{equation} 
where, as earlier, $u=\tau -\tau'$.

%%%%%%%%%%%%%%%%%%%%%%%%%%%%%%%%%%%%%%%%%%%%%%%%%%%%%%%%%%%%%%%%%%%%%%%%%%%%%%%%

\subsection{Odd statistics in odd dimensions for odd couplings}

Let me now assume that the quantum scalar field~${\hat \phi}$ 
is in the Minkowski vacuum.
In this case, the $(2\,n)$-point function $G_{n}\left({\tilde x}, 
{\tilde x'}\right)$ reduces to
\begin{equation}
G_{n}\left({\tilde x}, {\tilde x'}\right)
= n!\, \left[G^{+}\left({\tilde x}, {\tilde x'}\right)\right]^n,
\label{eq:2nptfn-mv}
\end{equation}
where $G^{+}\left({\tilde x}, {\tilde x'}\right)$ denotes the 
Wightman function in the Minkowski vacuum\footnote{I should stress
here that I would have arrived at the expression~(\ref{eq:2nptfn-mv}) 
for the $(2\,n)$-point function in the Minkowski vacuum even if I 
had started with the transition amplitude~(\ref{eq:ta-nlcd}) [instead 
of the normal ordered amplitude~(\ref{eq:no-ta-nlcd})], expressed the 
resulting $(2\,n)$-point function in the transition probability in 
terms of the two-point functions using Wick's theorem and then replaced 
the divergent terms that arise ({\it i.e.}\/ those two-point functions 
with coincident points) with the corresponding regularized 
expressions~\cite{suzuki-1997,sriram-2002b}.}.
The Wightman function~(\ref{eq:wfn-mv}) that I had quoted earlier
had corresponded to the result in $(3+1)$-spacetime dimensions. 
In $(D+1)$ spacetime dimensions [and for $(D+1)\ge 3$], the 
Wightman function for a massless scalar field in the Minkowski 
vacuum is given by~\cite{detectors-odd-std}
\begin{equation}
G^{+}({\tilde x},{\tilde x'})
= \frac{C_D}{\biggl\{(-1)\; \left[(t-t'-i\,\epsilon)^2
-\vert {\bm x}-{\bm x}'\vert^2\right]\biggr\}^{(D-1)/2}},
\label{eq:wfn-mv-ad}
\end{equation}
where it should be evident that ${\bm x}\equiv\left(x^{1},x^{2},
\ldots,x^{D}\right)$, while the quantity~$C_{D}$ is given by  
\begin{equation}
C_D= \Gamma\left[(D-1)/2\right]/\left[4\,\pi^{(D+1)/2}\right]
\label{eq:cd}
\end{equation}
with $\Gamma\left[(D-1)/2\right]$ denoting the Gamma function.

\par

Now, the trajectory of a detector accelerating uniformly along the 
$x^{1}$~direction with a proper acceleration~$g$ is given by 
\begin{equation}
{\tilde x}(\tau)=g^{-1}\, \left[{\rm sinh}\,(g\,\tau),\, 
{\rm cosh}\,(g\,\tau),\,0,\,0,\ldots,\,0\right],
\label{eq:uat-astd}
\end{equation}
where~$\tau$ is the proper time in the frame of the detector.
On substituting this trajectory in the Minkowski Wightman 
function~(\ref{eq:wfn-mv-ad}), I obtain 
that~\cite{detectors-odd-std}
\begin{equation}
G^{+}(u)
=\frac{C_{D}\; (g/2\,i)^{(D-1)}}{\biggl\{{\rm sinh}\left[
\left(g\,u/2\right)-i\,\epsilon\right]\biggr\}^{(D-1)}}.
\label{eq:wfn-mv-ad-uat}
\end{equation}
Therefore, along the trajectory of the uniformly 
accelerating detector, the $(2\,n)$-point function in 
the Minkowski vacuum~(\ref{eq:2nptfn-mv}) is given by
\begin{equation}
G_{n}(u)
=\frac{n!\, C_{D}^{n}\; 
(g/2\,i)^{p}}{\biggl\{{\rm sinh}\left[(g\,u/2)
-i\,\epsilon\right]\biggl\}^{p}},
\label{eq:2nptfn-mv-uat}
\end{equation}
where $p=(D-1)\,n$.

\par

Upon substituting the $(2\,n)$-point function~(\ref{eq:2nptfn-mv-uat}) 
in the expression~(\ref{eq:tpr-nlcd}) and carrying out the resulting 
integral~\cite{gradshteyn-1980}, I find that the transition probability 
rate of the uniformly accelerated, non-linearly coupled detector can 
be written as~\cite{sriram-2002b}
\begin{equation}
R_{n}(E) 
=B(n,D)\;\;
\left\{
\begin{array}{l}
\left(g^p/E\right)\;
{\underbrace{\frac{1}{\exp\,(2\,\pi\,E/g)-1}}}\;
\prod\limits_{l=0}^{(p-2)/2}
\left[l^2+(E/g)^2\right]\\
\qquad\quad\,\mbox{Bose-Einstein factor}\qquad
\qquad\qquad\quad\;\mbox{when $p$ is even},\\
g^{p-1}\;
{\underbrace{\frac{1}{\exp\,(2\,\pi\,E/g)+1}}}\;
\prod\limits_{l=0}^{(p-3)/2}
\biggl\{\left[(2\,l+1)/2\right]^2+(E/g)^2\biggr\}\\
\qquad\quad\;\mbox{Fermi-Dirac factor}\qquad
\qquad\qquad\qquad\mbox{when $p$ is odd,}
\end{array}\right.
\label{eq:tpr-nlcd-uat}
\end{equation}
where the quantity $B(n,D)$ is given by
\begin{equation}
B(n,D)=2\,\pi\,n!\; C_{D}^{n}/\Gamma(p).
\end{equation}
When $(D+1)$ is even, $p$ is even for all $n$ and, hence, 
a Bose-Einstein factor will always arise in the response of 
the uniformly accelerated detector in an even-dimensional flat 
spacetime.
Whereas, when $(D+1)$ is odd, evidently, $p$ will be odd or 
even depending on whether $n$ is odd or even.
Therefore, in an odd-dimensional flat spacetime, a Fermi-Dirac 
factor will arise in the detector response when~$n$ is odd (as
in the case of the Unruh-DeWitt detector), but a Bose-Einstein 
factor will appear when~$n$ is even!

\par

Let me make three clarifying comments regarding the curious result 
I have obtained above.
To begin with, the temperature associated with the Bose-Einstein 
and the Fermi-Dirac factors that appear in the response of the
non-linearly coupled detector is the standard Unruh temperature, 
{\it viz.}\/~$g/(2\,\pi)$. 
Moreover, the response of the detector is characterized 
{\it completely}\/ by either a Bose-Einstein or a Fermi-Dirac 
distribution {\it only}\/ in situations wherein $p<3$.
When $p\ge 3$, apart from a Bose-Einstein or a Fermi-Dirac factor,
the detector response contains a term which is polynomial in $E/g$.
Lastly, plots of the transition probability rate of the detector 
suggest that, though the characteristic response of the detector
alternates between the Bose-Einstein and the Fermi-Dirac factors 
as we go from one $D$ to another for odd $n$ [or from one $n$ to 
another when $(D+1)$ is odd], the complete spectra themselves 
exhibit a smooth dependence  on the index of non-linearity of 
the coupling as well as the dimension of spacetime (in this 
context, see the figures in Ref.~\cite{sriram-2002b}).

%%%%%%%%%%%%%%%%%%%%%%%%%%%%%%%%%%%%%%%%%%%%%%%%%%%%%%%%%%%%%%%%%%%%%%%%%%%%%%%%

\subsection{Nature of the odd statistics}

Despite its interesting character, the `inversion of statistics'
encountered in the response of the detector in odd dimensions
for odd couplings seems to be only {\it apparent}.\/
It is well known that, in the frame of the uniformly accelerating 
detector, the Wightman function in the Minkowski 
vacuum~(\ref{eq:wfn-mv-ad-uat}) is skew-periodic in imaginary proper 
time with a period corresponding to the inverse of the Unruh 
temperature~\cite{wfn-sp}, {\it i.e.}\/
\begin{equation} 
G^{+}(u)=G^{+}\left[-u+(2\,\pi\, i/g)\right].
\end{equation}
This property is known as the Kubo-Martin-Schwinger (KMS) condition,
as is applicable to scalar fields. 
Note that the above property is, in fact, satisfied by the Minkowski
Wightman function in {\it all}\/ 
dimensions~\cite{detectors-odd-std}.
Since the $(2\,n)$-point function in the Minkowski vacuum is 
proportional to the $n$th power of the Wightman function, obviously, 
in the frame of the accelerated detector, the $(2\,n)$-point function 
will also be skew-periodic in imaginary proper time for all $n$ and 
$D$ [cf. Eq.~(\ref{eq:2nptfn-mv-uat})]. 
In other words, the $(2\,n)$-point function satisfies the KMS condition 
(as is required for a scalar field) for {\it all}\/ $D$ and $n$.
This implies that the appearance of the Fermi-Dirac factor (instead of 
the expected Bose-Einstein factor) for odd $(D+1)$ and $n$ simply reflects 
a peculiar aspect of the detector rather than indicate a fundamental 
shift in the field theory in such situations~\cite{detectors-odd-std,
sriram-2002b,sriram-2003}.

%%%%%%%%%%%%%%%%%%%%%%%%%%%%%%%%%%%%%%%%%%%%%%%%%%%%%%%%%%%%%%%%%%%%%%%%%%%%%%%%

\section{Detecting Planck scale effects}\label{sec:pse}

Consider a typical mode that constitutes Hawking radiation at future 
null infinity around a collapsing black hole.
As one traces such a mode back to the past null infinity where the
initial conditions are imposed on the quantum field, it is found 
that the energy of the mode turns out to be way beyond the Planck 
scale~\cite{tpp-bhe}.
(This feature seems to have been originally noticed in
Ref.~\cite{wald-1976}; in this context, also see
Ref.~\cite{wald-1984}.) 
In fact, due to the rapid, virtually exponential expansion, a similar 
phenomenon is encountered in the context of the inflationary scenario.
One finds that scales of cosmological interest can be comparable to the
Planck scale at very early times when the initial conditions are imposed
during inflation~\cite{tpp-ic}.  
While the possible Planck scale corrections to Hawking radiation and 
the perturbations generated during inflation have cornered most of 
the attention~\cite{tpp-bhe,tpp-ic}, the Planck scale effects on a 
variety of non-perturbative, quantum field theoretic effects in flat 
as well as curved spacetimes have been investigated as well (see, for 
example, Refs.~\cite{srini-1998,jacobson-2001,other-e}).
In the absence of a viable quantum theory of gravity, it becomes 
imperative to extend such phenomenological analyses to as many
physical situations as possible (in this context, see 
Ref.~\cite{hossenfelder-2009}, and references therein).

\par

The Unruh effect has certain similarities with Hawking radiation from 
black holes.
Due to this reason, the Unruh effect and its variants provide another 
interesting domain to study the quantum gravitational effects~\cite{psm-ue}.
But, due to the lack of a workable quantum theory of gravity, to 
investigate the Planck scale effects, one is forced to consider 
phenomenological models constructed by hand.
These models attempt to capture one or more features expected of the 
actual effective theory obtained by integrating out the gravitational 
degrees of freedom. 
The approach based on modified dispersion relations has been extensively 
considered both in the context of black holes and inflationary cosmology.
In this approach, a fundamental scale is effectively introduced into the
theory by breaking local Lorentz invariance (see, for instance, 
Refs.~\cite{jacobson-2001,lvm-reviews}).
It should be clarified that there does not exist any experimental or 
observational reason to believe that Lorentz invariance could be 
violated at high energies.
Nevertheless, theoretically, these models prove to be attractive
because of the fact that they permit quantum field theories to be 
constructed and calculations to be carried out in a consistent 
fashion.

\par

In this section, I shall adopt the approach due to the modified dispersion 
relations to analyze the Planck scale corrections to the response of the 
rotating Unruh-DeWitt detector in flat spacetime.
As I shall show, the rotating trajectory turns out to be a special case 
wherein the transition probability rate of the rotating detector can be 
defined in precisely the same fashion as I had done earlier in the 
case of the canonical scalar field governed by the linear dispersion 
relation.
I shall illustrate that the response of the rotating detector can be 
computed {\it exactly},\/ although, numerically, even when the field it 
is coupled to is described by a non-linear dispersion relation. 

%%%%%%%%%%%%%%%%%%%%%%%%%%%%%%%%%%%%%%%%%%%%%%%%%%%%%%%%%%%%%%%%%%%%%%%%%%%%%%%%

\subsection{Scalar field governed by a modified dispersion 
relation}\label{sec:rd-wmdr}

I shall be interested in calculating the response of the rotating 
detector when it is coupled to a massless scalar field that is 
governed by a modified dispersion relation of the following form:
\begin{equation}
\omega=k\, \left[1+a\, \left(k/k_{_{\rm P}}\right)^{2}\right]^{1/2}.
\label{eq:mdr}
\end{equation}
The quantity $\omega$ is the frequency corresponding to the mode ${\bm k}$, 
$k=\vert {\bm k}\vert$ and $k_{_{\rm P}}$ denotes the fundamental scale
(that I shall assume to be of the order of the Planck scale) at which the 
deviations from the linear dispersion relation become important. 
Note that $a$ is a dimensionless constant whose magnitude is of order unity, 
and the above dispersion relation is super-luminal or sub-luminal depending 
upon whether $a$ is positive or negative.
Clearly, if I can evaluate the Wightman function associated with 
the quantized scalar field described by the non-linear dispersion 
relation~(\ref{eq:mdr}), I may then be able to evaluate the 
corresponding transition probability rate of the rotating detector 
as I had carried out originally. 
However, unlike the standard case, it turns out to be difficult to 
even arrive at an analytical expression for the Wightman function 
of such a scalar field.
Therefore, I shall make use of the second method that I had adopted 
earlier to evaluate the response of the rotating detector---I shall 
first integrate over the differential proper time and then numerically
sum over the normal modes to arrive at the transition probability rate.

\par

The equation of motion of the scalar field $\phi$ that is described by 
the dispersion relation~(\ref{eq:mdr}) is given by
\begin{equation}
\Box\,\phi
+\frac{a}{k_{_{\rm P}}^2}\, \nabla^2\,\left(\nabla^2\,\phi\right)=0,
\end{equation}
where $\Box$ is the d'Alembertian corresponding to the four dimensional
Minkowski spacetime, while $\nabla^2$ is the three dimensional, spatial 
Laplacian.
Evidently, the first term in the above equation is the standard one.
The non-linear term in the dispersion relation is responsible for the 
second term.
Such terms can be generated by adding suitable terms to the original 
action describing the scalar field~\cite{jacobson-2001,lvm-reviews}.
While these additional terms preserve rotational invariance, they break 
Lorentz invariance.
In fact, this property is common to all the theories that are described 
by a non-linear dispersion relation.
It is obvious that the normal modes of such a scalar field in flat 
spacetime remain plane waves as in the standard case, but with the 
frequency and the wavenumber related by the modified dispersion 
relation.
Moreover, the quantization of the scalar field can be carried out 
in the same fashion.
It is straightforward to show that, in the Minkowski vacuum, the 
Wightman function for any such field in $(3+1)$-spacetime
dimensions can be expressed as (see, for example, Ref.~\cite{jacobson-2001})
\begin{equation}
G^{+}_{_{\rm M}}({\tilde x},{\tilde x}')
=\int \frac{{\rm d}^{3}{\bm k}}{(2\,\pi)^3\, 2\, \omega}\, 
{\rm e}^{-i\,\omega\, (t-t')}\, 
{\rm e}^{i\,{\bm k}\cdot\, ({\bm x}-{\bm x}')}
\label{eq:mgfn-mdr}
\end{equation}
with $\omega$ being related to $k=\vert {\bm k}\vert$ by the given 
non-linear dispersion relation.

%%%%%%%%%%%%%%%%%%%%%%%%%%%%%%%%%%%%%%%%%%%%%%%%%%%%%%%%%%%%%%%%%%%%%%%%%%%%%%%

\subsection{Response of the rotating detector}

For a scalar field governed by a modified dispersion relation, using the 
expression~(\ref{eq:mgfn-mdr}) for the corresponding Wightman function, 
one can immediately show that, along the rotating trajectory, the function 
can be expressed exactly as in Eq.~(\ref{eq:mgfn-rt-ms}), with the 
frequency $\omega$ being related to the wavenumbers $q$ and $k_{z}$ by the 
non-linear dispersion relation.
Clearly, in such a case, the transition probability rate of the detector 
will again be given by Eq.~(\ref{eq:tpr-sc-rt-msev}) with $\omega$ suitably 
defined.
It is important to recognize that the result is actually applicable for
{\it any}\/ non-linear dispersion relation~\cite{gutti-2011}.

\par

Let me now evaluate the response of the rotating detector for the 
dispersion relation~(\ref{eq:mdr}).
In such a case, $\omega$ is related to the wavenumbers $q$ and $k_{z}$ 
as follows:
\begin{equation}
\omega
=\left(q^{2}+k_{z}^{2}\right)^{1/2}\; 
\left[1 + \frac{a}{k_{_{\rm P}}^2}\,
\left(q^{2}+k_{z}^{2}\right)\right] ^{1/2}.
\end{equation}
Also, one can show that the roots $\kappa_{z}$ [from Eq.~(\ref{eq:r-eq})] 
are given by
\begin{equation}
\kappa_{z}^2= \pm \frac{k_{_{\rm P}}^2}{2\,a}\,
\left(1+\frac{4\,a\,\lambda^2}{k_{_{\rm P}}^2}\right)^{1/2}
-\frac{k_{_{\rm P}}^2}{2\,a}-q^2,
\end{equation}
with $\lambda$ defined as in Eq.~(\ref{eq:delta}).
It ought to be noted that $\kappa_{z}^{2}$ has to be positive definite, 
since $\kappa_{z}$ is a real quantity. 

\par

Let me first consider the super-luminal case when $a$ is positive.
When, say, $a=1$, the two roots that contribute to the delta 
function in Eq.~(\ref{eq:tpr-sc-rt-msev}) can be written as
\begin{equation}
\kappa_{z}=\pm\left(\lambda_{+}^{2}-q^2\right)^{1/2},
\end{equation}
where $\lambda_{+}^{2}$ is given by the expression
\begin{eqnarray}
\lambda_{+}^{2}
&=&\frac{k_{_{\rm P}}^2}{2}\, 
\left[\left(1+\frac{4\,\lambda^2}{k_{_{\rm P}}^2}\right)^{1/2}
-1\right]\nonumber\\
&=&\frac{{\bar \lambda}_{+}^{2}}{\sigma^{2}}
=\frac{{{\bar k}_{_{\rm P}}}^2}{2\, \sigma^{2}}\, 
\left[\left(1+\frac{4\,(\sigma\,\Omega\, 
{\bar \lambda})^2}{{\bar k}_{_{\rm P}}^2}\right)^{1/2}-1\right].
\label{eq:lambda-plus}
\end{eqnarray}
Note that ${\bar k}_{_{\rm P}}=\sigma\, k_{_{\rm P}}$ denotes the 
dimensionless fundamental scale and the sub-script in $\lambda_{+}$ 
refers to the fact that I am considering a super-luminal dispersion 
relation.
Further, as $\kappa_{z}$ is real, I require that $q\le  \lambda_{+}$.
As in the standard case, the positive and negative roots of 
$\kappa_{z}$ above contribute equally.
Therefore, the response of the rotating detector is given by
\begin{equation}
{\bar R}({\bar E})
=\sigma\, R({\bar E})
=\frac{\sigma}{2\,\pi\,\gamma}\,
\sum^{\infty}_{m\ge {\bar E}}\, 
\left(1+\frac{2\,\lambda_{+}^2}{k_{_{\rm P}}^2}\right)^{-1}\;
\int\limits_{0}^{\lambda_{+}}\; {\rm d}q\; q\;
\left[\frac{J_{m}^{2}(q\,\sigma)}{\left(\lambda_{+}^{2}
-q^{2}\right)^{1/2}}\right],
\end{equation}
and the integral over $q$ can be carried out as in the standard case  
to arrive at the result
\begin{eqnarray}
{\bar R}({\bar E})
&=&\frac{1}{2\,\pi\,\gamma}\
\sum^{\infty}_{m\ge {\bar E}}\,
\frac{{\bar \lambda}_{+}^{(2\,m+1)}}{\Gamma\,\left(2\,m+2\right)}\; 
\left(1+\frac{2\,{{\bar \lambda}_{+}}^2}{{\bar k}_{_{\rm P}}^2}
\right)^{-1}\nonumber\\
& &\times\,_1F_2\left[m+(1/2);\,m+(3/2),2\,m+1;-{\bar \lambda}_{+}^2\right].
\label{eq:tpr-rd-psm-spldr}
\end{eqnarray}
It should be emphasized here that this result for the transition 
probability rate is exact and no approximations have been made in 
arriving at the expression.

\par

Since the Planck scale is expected to be orders of magnitude beyond
the scales probed by experiments, the quantity ${\bar k}_{_{\rm P}}$ 
is expected to be large.
It is clear that, as ${\bar k}_{_{\rm P}}\to \infty$, ${\bar \lambda}_{+}
\to \sigma\, \Omega\, {\bar \lambda}$ and, hence, the transition 
transition probability rate~(\ref{eq:tpr-rd-psm-spldr}) reduces 
to the expression that I had arrived at earlier for the standard
dispersion relation [{\it viz.}\/~Eq.~(\ref{eq:tpr-sc-rt-smfv})], 
as required.
Let me now evaluate the Planck scale corrections to the standard result 
by expanding the transition probability rate~(\ref{eq:tpr-rd-psm-spldr})
in terms of $\lambda/k_{_{\rm P}}$ and retaining terms upto 
${\cal O}[\left(\lambda/k_{_{\rm P}}\right)^{2}]$. 
Note that, in such a case, $\lambda_{+}$ reduces to
\begin{equation}
\lambda_{+} 
\simeq
\lambda\, \left(1 - \frac{\lambda^{2}}{2\,k_{_{\rm P}}^{2}}\right),
\end{equation} 
so that I have
\begin{equation}
\lambda_+^{(2\,m+1)} 
\simeq \lambda^{(2\,m+1)}
-(2\,m+1)\, \frac{\lambda^{(2\, m+3)}}{2\, k_{_{\rm P}}^2}
\end{equation}
and
\begin{equation}
\left(1+\frac{2\,\lambda_{+}^2}{k_{_{\rm P}}^2}\right)^{-1} 
\simeq 1-\frac{2\,\lambda^2}{k_{_{\rm P}}^2}.
\end{equation}
Moreover, in the limit of our interest, the hypergeometric function 
in Eq.~(\ref{eq:tpr-rd-psm-spldr}) can be written as 
\begin{eqnarray}
& &\!\!\!\!\!\!\!\!\!\!
_1F_2\left[m+(1/2);\,m+(3/2),\,2\,m+1;\,-{\bar \lambda}_{+}^2\right]\nonumber\\
& &\quad\simeq\, 
_1F_2\left[m+(1/2);\,m+(3/2),\,2\,m+1;\,
-\left(\sigma\,\Omega\, {\bar \lambda}\right)^2\right]\nonumber\\
& &\qquad+\, \frac{(\sigma\, \Omega\,{\bar \lambda})^{2}}{{\bar k}_{_{\rm P}}^2}\,
\frac{[m+(1/2)]\; (\sigma\, \Omega\,{\bar \lambda})^{2}}{[m+(3/2)]\; 
(2\, m+1)}\nonumber\\
& &\qquad\;\;
\times\,_1F_2\left[m+(3/2);\,m+(5/2),\,2\,m +2;\,
-\left(\sigma\,\Omega\, {\bar \lambda}\right)^2\right].
\end{eqnarray}
Upon using the above expansions, I obtain the response of the 
detector at ${\cal O}[\left(\lambda/k_{_{\rm P}}\right)^{2}]$ to be
\begin{eqnarray}
\!\!\!\!\!\!\!\!\!\!\!\!\!\!\!\!\!\!\!\!\!\!\!\!
{\bar R}({\bar E})
&\simeq&\frac{1}{2\,\pi\,\gamma}\,
\sum^{\infty}_{m\ge {\bar E}}\,
\frac{\left(\sigma\,\Omega\, 
{\bar \lambda}\right)^{(2\,m+1)}}{\Gamma\left(2\,m+2\right)}\nonumber\\
& &\times\,_1F_2\left[m+(1/2);\,m+(3/2),\,(2\,m+1);\,
-\left(\sigma\,\Omega\, {\bar \lambda}\right)^2\right]\nonumber\\
& & - \, \frac{1}{2\,\pi\,\gamma}\,
\frac{(\sigma\, \Omega\,{\bar \lambda})^{2}}{{\bar k}_{_{\rm P}}^2}\,
\sum^{\infty}_{m\ge {\bar {\cal E}}}\,
\frac{[m+(5/2)]\,
\left(\sigma\,\Omega\,{\bar \lambda}\right)^{(2\,m+1)}}{\Gamma\left(2\, 
m+2\right)}\nonumber\\
& &\times\, _1F_2\left[m+(1/2);\,m+(3/2),\, 2\,m+1;\,
-\left(\sigma\,\Omega\,{\bar \lambda}\right)^2\right]\nonumber\\
& & + \, \frac{1}{2\,\pi\,\gamma}\,
\frac{(\sigma\, \Omega\,{\bar \lambda})^{2}}{{\bar k}_{_{\rm P}}^2}\,
\sum^{\infty}_{m\ge {\bar E}}\,
\frac{[m + (1/2)]\; \left(\sigma\,\Omega\,
{\bar \lambda}\right)^{(2\,m+3)}}{[m+(3/2)]\, (2\,m+1)\;
\Gamma\left(2\,m+2\right)}\nonumber\\
& &\times\;
 _1F_2\left[m+(3/2);\,m+(5/2),\,2\,m+2;\,
-\left(\sigma\,\Omega\,{\bar \lambda}\right)^2\right]. 
\end{eqnarray}
Evidently, the first term in this expression corresponds to the
conventional transition probability rate 
[cf. Eq.~(\ref{eq:tpr-sc-rt-smfv})], while the other two terms 
represent the leading corrections to the standard result.

\par
  
Let me now turn to considering the sub-luminal dispersion relation. 
When $a$ is negative, say, $a=-1$, the roots $\kappa_{z}$ are given by
\begin{equation}
\kappa_{z}=\pm\left(\lambda_{-}^{2}-q^2\right)^{1/2}
\end{equation}
with $\lambda_{-}^{2}$ defined as
\begin{eqnarray}
(\lambda_{-}^{\pm})^{2}
&=&\frac{k_{_{\rm P}}^2}{2}\, 
\left[1\pm\left(1-\frac{4\,\lambda^2}{k_{_{\rm P}}^2}\right)^{1/2}\right]\nonumber\\
&=&\frac{({\bar \lambda}_{-}^{\pm})^{2}}{\sigma^{2}}
=\frac{{\bar k}_{_{\rm P}}^2}{2\, \sigma^2}\, 
\Biggl\{1\pm\left[1-\frac{4\,(\sigma\, \Omega\,
{\bar \lambda})^2}{{\bar k}_{_{\rm P}}^2}\right]^{1/2}\Biggr\},
\end{eqnarray}
where the minus sign in the sub-script represents that it corresponds 
to the sub-luminal case ({\it i.e.}\/ when $a$ is negative), while the 
super-scripts denote the two different possibilities of $\lambda_{-}$.  
Just as in the super-luminal case ({\it i.e.}\/ when $a=1$), I require 
$q\le \lambda_{-}^{\pm}$, if $\kappa_{z}$ is to remain real.
Moreover, note that, unlike the super-luminal case, there also arises
an upper limit on the sum over $m$.
I require that $\lambda\leq k_{_{\rm P}}/2$, in order to ensure that 
$\lambda_{-}^{\pm}$ is real.
This corresponds to $m\le {\bar E}+{\bar k}_{_{\rm P}}/(2\,\sigma\, \Omega)$.
Therefore, for the sub-luminal dispersion relation, I find that I 
can write the response of the rotating detector as follows:
\begin{eqnarray}
{\bar R}({\bar E})
&=&\frac{1}{2\,\pi\,\gamma}\,
\sum^{{\bar E}+{\bar k}_{_{\rm P}}/(2\,\sigma\,\Omega)}_{m\ge {\bar E}}\,
\frac{\left({\bar \lambda}_{-}^{-}\right)^{(2\,m+1)}}{\Gamma(2\,m+2)}\,
\left(\;\left\vert\, 1
-\frac{2\,({\bar \lambda}_{-}^{-})^2}{{\bar k}_{_{\rm P}}^2}\right
\vert\;\right)^{-1}\nonumber\\ 
& &\times\, _1F_2\left[m+(1/2);\,m+(3/2),\,2\,m+1;\,
-\left({\bar \lambda}_{-}^{-}\right)^2\right]\nonumber\\
& &+\,\frac{1}{2\,\pi\,\gamma}\,
\sum^{{\bar E}+k_{_{\rm P}}/(2\,\sigma\,\Omega)}_{m\ge {\bar E}}\,
\frac{\left({\bar \lambda}_{-}^{+}\right)^{2\,m+1}}{\Gamma\left(2\,m+2\right)}\,
\left(\;\left\vert\, 1-\frac{2\,({\bar \lambda}_{-}^{+})^2}{{\bar k}_{_{\rm P}}^2}
\right\vert\;\right)^{-1}\nonumber\\ 
& &\times\; _1F_2\left[m+(1/2);\,m+(3/2),\,2\,m+1;\,
-\left({\bar \lambda}_{-}^{+}\right)^2\right].
\label{eq:tpr-rd-psm-sbldr}
\end{eqnarray}

\par

The reason for the upper limit on $m$ as well as the origin of the second 
term in the above expression for the response of the rotating detector can 
be easily understood. 
The quantity $\omega$ is a monotonically increasing function of $q$ and 
$k_{z}$ in the case of the super-luminal dispersion relation.
Because of this reason, there exist only two real roots of $k_{z}$ 
corresponding to a given $\omega$. 
Moreover, $\omega^{2}$ remains positive definite for all the modes.
In contrast, in the sub-luminal case, after a rise, $\omega$ begins to 
decrease for sufficiently large values of $q$ and $k_{z}$.
Actually, $\omega^{2}$ even turns negative at a suitably large 
value~\cite{jacobson-2001}. 
It is this feature of the sub-luminal dispersion relation which leads to 
the upper limit on $m$, and the limit ensures that we avoid complex 
frequencies. 
(Such a cut-off can be achieved if I assume that, say, the detector is not 
coupled to modes with $m$ beyond a certain value, when the frequency turns 
complex.)
There arise two additional two roots of $k_{z}$ which contribute to the 
detector response in the sub-luminal case as a result of the decreasing 
$\omega$ at large $q$ and $k_{z}$.
The second term in the above transition probability rate of the rotating
detector corresponds to the contributions from these two extra roots.

\par

\begin{figure}[!t]
\begin{center}
\resizebox{285pt}{190pt}{\includegraphics{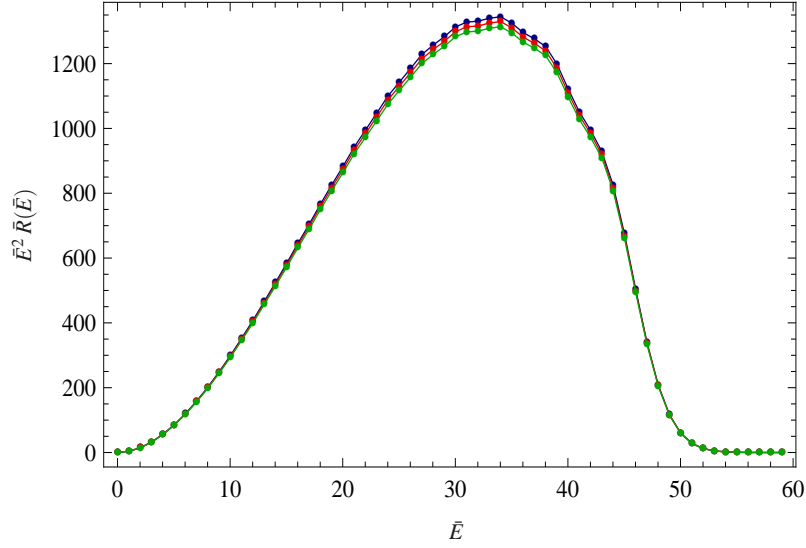}}
\vskip -120 pt \hskip -305 pt 
\rotatebox{90}{${\bar E}^2\, {\bar R}({\bar E})$}
\vskip 95 pt
\hskip 22 pt ${\bar E}$ 
\caption{The transition probability rate of the rotating Unruh-DeWitt 
detector that is coupled to a massless scalar field governed by the 
modified dispersion relation~(\ref{eq:mdr}) with $a=-1$.
The dots and the curves linking them denote the numerical results for
the same set of values for the quantity $\sigma\, \Omega$ (and the same
choice of colors) that I had plotted in the previous figure.
I have set ${\bar k}_{_{\rm P}}=50$, which is an extremely small value 
for ${\bar k}_{_{\rm P}}$.
Evidently, for such a value, the modifications to the standard result 
(cf. Fig.~\ref{fig:rd-sc}) due to the sub-luminal dispersion relation 
is considerable.
In fact, more realistic values of ${\bar k}_{_{\rm P}}$ would correspond
to, say, ${\bar k}_{_{\rm P}}/{\bar E}> 10^{10}$.
However, numerically, it turns out to be difficult to sum the contributions
in the expression~(\ref{eq:tpr-rd-psm-sbldr}) up to such large values of 
${\bar k}_{_{\rm P}}$.
It seems reasonable to conclude that the modifications to standard result 
due to the sub-luminal dispersion relation can be expected to be much 
smaller if one assumes ${\bar k}_{_{\rm P}}$ to be sufficiently large.
Nevertheless, my analysis unambiguously points to the fact that, as is 
known to occur in other situations, a sub-luminal dispersion relation  
modifies the standard result considerably more than a similar 
super-luminal dispersion relation.}\label{fig:rd-psm}
\end{center}
\end{figure}

If one plots the result~(\ref{eq:tpr-rd-psm-spldr}) for the response of 
the rotating detector when it is coupled to a field that is governed by 
a super-luminal dispersion relation, one finds that it does not differ 
from the standard result (as plotted in Fig.~\ref{fig:rd-sc}) even for 
an unnaturally small value of ${\bar k}_{_{\rm P}}$ such that, say, 
${\bar k}_{_{\rm P}}/{\bar E}\simeq 10$.
This implies that super-luminal dispersion relations do not alter the 
conventional result to any extent.
It needs to be emphasized here that similar conclusions have been arrived 
at earlier in the context of black holes as well as inflationary cosmology.
In these contexts, it has been shown that Hawking radiation and 
the inflationary perturbation spectra remain unaffected due to 
super-luminal modifications to the conventional, linear, dispersion 
relation~\cite{tpp-bhe,tpp-ic}.
In Fig.~\ref{fig:rd-psm}, I have plotted the transition probability 
rate~(\ref{eq:tpr-rd-psm-sbldr}) of the rotating Unruh-DeWitt detector
corresponding to the sub-luminal dispersion relation that I have
considered.
I have plotted the result for a rather small value of ${\bar k}_{_{\rm P}}=50$.
It is clear from the figure that the sub-luminal dispersion relation can 
lead to substantial modifications to the standard result.
I believe that the modifications from the standard result will be 
considerably smaller (than exhibited in the figure) for much larger 
and more realistic values of ${\bar k}_{_{\rm P}}$ such that, say, 
${\bar k}_{_{\rm P}}/{\bar E} > 10^{10}$.
%%%%%%%%%%%%%%%%%%%%%%%%%%%%%%%%%%%%%%%%%%%%%%%%%%%%%%%%%%%%%%%%%%%%%%%%%%%
%%%%%%%%%%%%%%%%%%%%%%%%%%%%%%%%%%%%%%%%%%%%%%%%%%%%%%%%%%%%%%%%%%%%%%%%%%%%%%%

\subsection{Rotating detector in the presence of a boundary}\label{sec:rd-wb}

I shall now consider an interesting situation wherein I study the 
response of the rotating detector in the presence of an additional 
boundary condition that is imposed on the scalar field on a 
cylindrical surface in flat spacetime.
Because of the symmetry of the problem, in this case too, the
cylindrical coordinates turn out to be more convenient to work with.

\par

It is well known that the time-like Killing vector associated with an 
observer who is rotating at an angular velocity $\Omega$ in flat 
spacetime becomes space-like for radii greater than $\rho_{_{\rm SL}}
=1/\Omega$.
Due to this reason, it has been argued that one needs to impose a boundary 
condition on the quantum field at a radius $\rho<\rho_{_{\rm SL}}$ when 
evaluating the response of a rotating detector~\cite{davies-1996}. 
Curiously, in the presence of such a boundary, it was found that a 
rotating Unruh-DeWitt detector which is coupled to the standard scalar 
field ceases to respond.
It is then interesting to examine whether this result holds true even
when one assumes that the scalar field is governed by a modified dispersion 
relation. 

\par

In the cylindrical coordinates, along the rotating
trajectory~(\ref{eq:rt-cyc}), the Wightman function corresponding to a 
scalar field that is assumed to vanish at, say, 
$\rho=\rho_{\ast}\;(<\rho_{_{\rm SL}})$, can be expressed as a sum over 
the normal modes of the field as follows~\cite{davies-1996}:
\begin{equation}
G^{+}(u)
=\sum\limits_{m=-\infty}^{\infty}\; \sum\limits_{n=1}^{\infty}\; 
\int\limits_{-\infty}^{\infty}\; \frac{{\rm d}k_{z}}{\left(2\,\pi\right)^{2}\,
2\,\omega}\;
\left[N\, J_{m}(\xi_{mn}\,\sigma/\rho_{\ast})\right]^{2}\; 
e^{-i\,\gamma\,\left(\omega-m\,\Omega\right)\, u},
\label{eq:cgfn-rt}
\end{equation}
where $\xi_{mn}$ denotes the $n$th zero of the Bessel function 
$J_{m}(\xi_{mn}\,\sigma/\rho_{\ast})$, while $N$ is a normalization 
constant that is given by
\begin{equation}
N=\frac{\sqrt{2}}{\rho_{\ast}\,\vert J_{m+1}(\xi_{mn})\vert}.
\end{equation}
As in the situation without a boundary, $m$ is a real integer, whereas 
$k_{z}$ is a continuous real number. 
But, due to the imposition of the boundary condition at $\rho=\rho_{\ast}$, 
the spectrum of the radial modes is now discrete, and is described by the
positive integer~$n$.
It should be pointed out that the expression~(\ref{eq:cgfn-rt}) is 
in fact valid for any dispersion relation, with $\omega$ suitably 
related to the quantities $\xi_{mn}$ and $k_{z}$.
For instance, in the case of the modified dispersion 
relation~(\ref{eq:mdr}), the quantity $\omega$ is given by
\begin{equation}
\omega
=\left(\frac{\xi_{mn}^2}{\rho_{\ast}^2} + k_{z}^2\right)^{1/2}\;
\left[1+\frac{a}{k_{_{\rm P}}^{2}}\;
\left(\frac{\xi_{mn}^2}{\rho_{\ast}^2}
+k_{z}^{2}\right)\right]^{1/2},\label{eq:wrc}
\end{equation}
where, it is evident that, while the overall factor corresponds to the 
standard, linear, dispersion relation, the term involving $a$
within the brackets arises due to the modifications to it.
Since the Wightman function depends only $u$, the transition probability 
rate of the detector simplifies to
\begin{equation}
R(E)
=\sum\limits_{m=-\infty}^{\infty}\; \sum\limits_{n=1}^{\infty}\;
\int\limits_{-\infty}^{\infty}\; \frac{{\rm d}k_{z}}{2\,\pi\;\,
2\,\omega}\; 
\left[N\, J_{m}(\xi_{mn}\,\sigma/\rho_{\ast})\right]^{2}\; 
\delta^{(1)}\left[E+\gamma\, \left(\omega-m\,\Omega\right)\right].
\end{equation}
For exactly the same reasons that I had presented in the last 
section, the delta function in this expression can be 
non-zero only when $m>0$.
In fact, the detector will respond only under the condition
\begin{equation}
m\,\Omega 
> \frac{\xi_{m1}}{\rho_{\ast}}\; \left(1 
+ \frac{a}{k_{_{\rm P}}^2}\, \frac{\xi_{m1}^2}{\rho_{\ast}^2}\right)^{1/2},
\end{equation}
where the right hand side is the lowest possible value of $\omega$ 
corresponding to $n=1$ and $k_{z}=0$.
However, from the properties of the Bessel function, it is known
that $\xi_{mn}>m$, for all $m$ and $n$ (see, for instance, 
Ref.~\cite{abramowitz-1965}).
Therefore, {\it when $a$ is positive},\/ $\Omega\, \rho_{\ast}$ has 
to be greater than unity, if the rotating detector has to respond. 
But, this is not possible since I have assumed that the boundary at 
$\rho_{\ast}$ is located {\it inside}\/ the static limit $\rho_{_{\rm SL}}
=1/\Omega$.
This is exactly the same conclusion that one arrives at in the standard 
case~\cite{davies-1996,crispino-2008}.

\par

Actually, it is easy to argue that the above conclusion would apply for 
all super-luminal dispersion relations.
But, it seems that, under the same conditions, the rotating detector 
would be excited by a certain range of modes if I consider the scalar 
field to be described by a sub-luminal (such as, when $a <0$) 
dispersion relation!
In fact, this aspect is rather easy to understand.
Consider a frequency, say, $\omega$, associated with a mode through the
linear dispersion relation. 
Evidently, a super-luminal dispersion relation raises the energy of all
the modes, while the sub-luminal dispersion relation lowers it.
Therefore, if the interaction of the detector with a standard field does 
not excite a particular mode of the quantum field, clearly, the mode 
is unlikely to be excited if its energy has been raised further, as in a 
super-luminal dispersion relation.
However, the motion of the detector mode may be able to excite a mode 
of the field, if the energy of certain modes are lowered when compared 
to the standard case, as the sub-luminal dispersion relation does.

%%%%%%%%%%%%%%%%%%%%%%%%%%%%%%%%%%%%%%%%%%%%%%%%%%%%%%%%%%%%%%%%%%%%%%%%%%%%%%%%

\section{Finite time detectors}\label{sec:ftd}

The response of detectors have always been studied for their entire 
history, {\it viz.}\/ from the infinite past to the infinite future in the 
detector's proper time. 
But, in any realistic situation, the detectors can be kept switched 
on only for a finite period of time and due to this reason the study 
of the response of a detector for a finite interval in proper time 
becomes important. 
In this section, I shall illustrate that, unless the detectors are 
switched on smoothly, the response of the detector can contain divergent
contributions~\cite{detectors-ft,sriram-1996}. 
 
\par

Consider a Unruh-DeWitt detector that has been switched on for a 
finite period of time with the aid of a window function, say, 
$W(\tau,T)$, where, as before, $\tau$ is the proper time in the 
frame of the detector, while $T$ is the effective time for which 
the detector is turned on.
The window function $W(\tau, T)$ can be expected to have the following
properties: 
\begin{equation}
W(\tau, T)
\simeq 
\left\{\begin{array}{ll}
1 & {\rm for}~\vert \tau \vert \ll T,\\
0 & {\rm for}~\vert \tau \vert \gg T.
\end{array}\right.
\end{equation}
In such a case, instead of Eq.~(\ref{eq:tp}), the transition probability 
of the detector will be described by the integral
\begin{equation}
P(E, T) = \int_{-\infty}^\infty {\rm d}\tau\, 
\int_{-\infty}^\infty {\rm d}\tau'\, {\rm e}^{-i\, E\, (\tau-\tau')}\,
W(\tau, T)\, W(\tau', T)\, 
G^{+}\left[{\tilde x}(\tau), {\tilde x}(\tau')\right].
\end{equation}
While abrupt switching corresponds to 
\begin{equation}
W(\tau, T) = \Theta(T-\tau) + \Theta(T+\tau), 
\end{equation}
more gradual switching on and off can be achieved, for instance, with 
the aid of the window function
\begin{equation}
W(\tau, T) 
= \exp-\left(\frac{\tau^2}{2\,T^2}\right).
\end{equation}

\par

Consider a detector that is moving along the integral curve of a 
time-like Killing vector field so that $G^{+}\left[{\tilde x}(\tau), 
{\tilde x}(\tau')\right]=G^{+}(\tau-\tau')$.
Let the detector be switched on and off with the aid of a smooth 
window function of the form~$W(\tau/T)$. 
In such a situation, I can express the transition probability of the 
detector as
\begin{eqnarray}
P(E, T) 
&=& \int_{-\infty}^{\infty} {\rm d}\tau\, 
\int_{-\infty}^{\infty}{\rm d}\tau'\, W(\tau, T)\, W(\tau', T)\, 
{\rm e}^{-i\, E\, (\tau-\tau')}\, G^{+}(\tau-\tau')\\
&=& W\left(i\,\frac{\partial}{\partial E}, T\right)\, 
W\left(-i\,\frac{\partial}{\partial E}, T\right)\, P(E),
\end{eqnarray}
where $P(E)$ is the original transition probability~(\ref{eq:tp}) for
the case of the Unruh-DeWitt detector that has been kept on for its 
entire history. 
Let me now expand $W({\tau, T})= W({\tau/ T})$ as a Taylor series around 
$\tau=0$ and assume that $W(0) = 1$, $W'(0) = 0$, where the overprime
denotes differentiation with respect to the argument $\tau/T$.
I can then write the window function as
\begin{eqnarray}
W\left(\frac{\tau}{T}\right)  
&\simeq& W(0) + W'(0)\, \left(\frac{\tau}{T}\right)
+\frac{1}{2}\, W''(0)\, {\left(\frac{\tau}{T}\right)}^2\nonumber\\ 
&\simeq& 1 + \frac{1}{2}\,W''(0)\,{\left(\frac{\tau}{T}\right)}^{2},
\end{eqnarray}
so that the transition probability becomes
\begin{eqnarray}
P(E,T) 
&\simeq& \left(1 - \frac{W''(0)}{2\, T^2}\,
\frac{{\partial}^2 }{\partial E^2}\right)^2\, P(E)\nonumber\\
&\simeq& P(E) - \frac{W''(0)}{T^2}\;
\frac{\partial^2 P(E)}{\partial E^2}.
\end{eqnarray}
This gives the transition probability rate to be
\begin{equation}
R(E, T) 
= R(E) - \frac{W''(0)}{T^2}\, 
\frac{{\partial}^2 R(E)}{\partial E^2}
+ {\cal O}\left(\frac{1}{T^4}\right),
\end{equation}                                
for any window function and trajectory. 
Note that the response at finite~$T$ depends on the derivatives of the 
window function, such as, for example, $W''(0)$. 
Hence, if the detector is switched on abruptly, these derivatives 
can diverge, thereby leading to divergent responses~\cite{detectors-ft}.

%%%%%%%%%%%%%%%%%%%%%%%%%%%%%%%%%%%%%%%%%%%%%%%%%%%%%%%%%%%%%%%%%%%%%%%%%%%%%%%%

\section{Summary}\label{sec:s}

The concept of detectors was originally introduced to provide an 
operational definition to the concept of a particle.
With this aim, the response of detectors have been studied in the 
literature in a wide variety of situations. 
In this article, I have described a few different aspects of 
detectors. 
I have highlighted the point that, while the detectors are sensitive 
to the phenomenon of particle production, their response do not, in 
general, reflect the particle content of the field.
I have shown that, in odd spacetime dimensions, the response of a 
detector that is coupled to an odd power of the scalar field exhibits 
a Fermi-Dirac distribution rather than the expected Bose-Einstein
distribution. 
I have also discussed the response of a rotating detector that is 
coupled to a scalar field governed by modified dispersion relations,
supposedly arising due to quantum gravitational effects.
I have illustrated that, as it has been encountered in other similar
contexts, while super-luminal dispersion relations hardly affect 
the response of the detector, sub-luminal relations substantially
modify the response.
Finally, I have argued that detectors which are switched on
abruptly can exhibit responses which contain divergences. 

%%%%%%%%%%%%%%%%%%%%%%%%%%%%%%%%%%%%%%%%%%%%%%%%%%%%%%%%%%%%%%%%%%%%%%%%%%%%%%%%

\begin{acknowledgement}
It is a pleasure to contribute an article to the volume being put together 
to celebrate the sixtieth birthday of Prof.~T.~Padmanabhan, or Paddy as he 
is affectionately known.
With undiminished energy and enthusiasm, Paddy continues to be an inspiration
for many of us.
I can not thank him adequately enough for the years of constant friendship, 
support and guidance. 
\end{acknowledgement}

%%%%%%%%%%%%%%%%%%%%%%%%%%%%%%%%%%%%%%%%%%%%%%%%%%%%%%%%%%%%%%%%%%%%%%%%%%%%%%%%

\end{document}